\RequirePackage{fix-cm}
\documentclass[10pt,twocolumn,english,reprint]{revtex4-1}
\usepackage[T1]{fontenc}
\usepackage[latin9]{inputenc}
\usepackage{geometry}
\usepackage{fancyhdr}
\usepackage{color}
\usepackage{babel}
\usepackage{amsmath}
\usepackage{amssymb}
\usepackage{graphicx}
\usepackage{esint}
\usepackage[unicode=true]
 {hyperref}

\makeatletter

\@ifundefined{textcolor}{}
{%
 \definecolor{BLACK}{gray}{0}
 \definecolor{WHITE}{gray}{1}
 \definecolor{RED}{rgb}{1,0,0}
 \definecolor{GREEN}{rgb}{0,1,0}
 \definecolor{BLUE}{rgb}{0,0,1}
 \definecolor{CYAN}{cmyk}{1,0,0,0}
 \definecolor{MAGENTA}{cmyk}{0,1,0,0}
 \definecolor{YELLOW}{cmyk}{0,0,1,0}
}

\makeatother

\begin{document}

\title{ }

\title{}

\title{Broken boost invariance in the Glasma via finite nuclei thickness}

\author{Andreas Ipp, David M\"{u}ller}

\address{\emph{Institut f\"{u}r Theoretische Physik, Technische Universit\"{a}t
Wien, }\\
\emph{Wiedner Hauptstr. 8-10, A-1040 Vienna, Austria}\\
\emph{E-Mail}: \href{mailto:ipp@hep.itp.tuwien.ac.at}{ipp@hep.itp.tuwien.ac.at},
\href{mailto:david.mueller@tuwien.ac.at}{david.mueller@tuwien.ac.at}}
\begin{abstract}
We simulate the creation and evolution of non-boost-invariant Glasma
in the early stages of heavy ion collisions within the color glass
condensate framework. This is accomplished by extending the McLerran-Venugopalan
model to include a parameter for the Lorentz-contracted but finite
width of the nucleus in the beam direction. We determine the rapidity
profile of the Glasma energy density, which shows deviations from
the boost-invariant result. Varying the parameters both broad and
narrow profiles can be produced. We compare our results to experimental
data from RHIC and find surprising agreement. 
\end{abstract}
\maketitle

\global\long\def\t{\dagger}
\global\long\def\p{\partial}
\global\long\def\ev#1{\left\langle #1\right\rangle }
\global\long\def\e{\varepsilon}

Heavy ion collisions at the Relativistic Heavy Ion Collider (RHIC)
and the Large Hadron Collider (LHC) provide insight into the properties
of nuclear matter under extreme conditions. The evolution of the Quark-Gluon
Plasma (QGP) that is created in such collisions is well described
by relativistic viscous hydrodynamics \citep{Heinz:2013th,Gale:2013da}.
A first principles description of the initial state of heavy-ion
collisions is provided by the Color Glass Condensate (CGC) framework
\citep{Gelis:2010nm,Gelis:2012ri,Gelis:2016rnt}. The CGC is a classical
effective field theory for nuclear matter at ultrarelativistic energies.
Models such as the IP-Glasma \citep{Schenke:2012wb,Schenke:2012fw}
in combination with hydrodynamics are able to correctly reproduce
azimuthal anisotropies and event-by-event multiplicity distributions
\citep{Gale:2012rq,Snellings:2011sz}. Furthermore, the CGC can explain
long-range rapidity correlations like the ridge \citep{Dumitru:2008wn,Iancu:2012xa}.

A Gaussian shaped rapidity profile of particle multiplicity can be
found in experiments covering various energy ranges, from LHC \citep{Abbas:2013bpa}
to RHIC Beam Energy Scan \citep{Bearden:2004yx,Flores:2016mtp}. This
shape is well explained by the Landau model \citep{Landau:1953gs}
up to RHIC energies \citep{Bearden:2004yx}, which assumes full stopping
of the colliding nuclei. The Landau model is in contrast to the Bjorken
model \citep{Bjorken:1982qr} which relies on approximate boost invariance.
A Gaussian profile has also been found in holographic calculations
of colliding shock waves \citep{Casalderrey-Solana:2013aba,vanderSchee:2014qwa,vanderSchee:2015rta}. 

In its original formulation collisions in the CGC picture are assumed
to be boost-invariant \citep{Kovner:1995ja,Kovner:1995ts,Krasnitz:1998ns,Lappi:2011ju}
and were thus only understood as an approximation valid close to midrapidity.
 This approach implicitly assumes infinitely thin Lorentz-contracted
nuclei and entails a classical, boost-invariant evolution of the Glasma
at leading order. 
\begin{figure}
\begin{centering}
\includegraphics{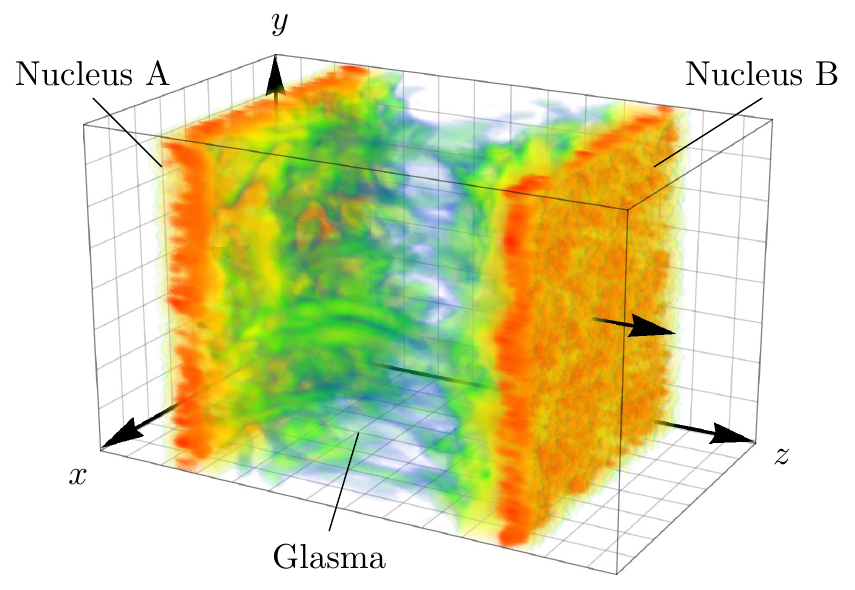}
\par\end{centering}

\protect\caption{A 3+1 dimensional colored particle-in-cell simulation of the collision
of two thick sheets of relativistic nuclear matter. The simulation
box covers a small part of the full transverse extent of the nuclei
in the $x,y$ plane. This figure shows a density plot of the energy
density of both nuclei A and B, and the three-dimensional Glasma that
is created in the collision.\label{fig:CPIC}}
\end{figure}

Only at the next-to-leading order the boost invariance is broken by
a change of the initial conditions through JIMWLK evolution \citep{JalilianMarian:1996xn,Iancu:2000hn,Mueller:2001uk,Ferreiro:2001qy},
and non-boost-invariant rapidity profiles can be obtained.\textcolor{red}{{}
}Recently such rapidity dependencies have been found to agree reasonably
well with experimental data where observables like charged particle
multiplicities show a Gaussian rapidity profile \citep{Schenke:2013dpa,Schenke:2016ksl}.
On the other hand it has been suggested that if one considers nuclei
with finite extent in the beam direction, deviations from boost invariance
may arise already at the classical level \citep{Ozonder:2013moa}.
In the case of proton-nucleus collisions methods have been developed
to systematically include finite width corrections of the nucleus
\citep{Altinoluk:2014oxa,Altinoluk:2015gia,Altinoluk:2015xuy}. However,
so far there has been no consistent simulation of the subsequent three-dimensional
evolution for heavy-ion collisions even at the classical level. 

In this letter, we show that Gaussian rapidity profiles of energy
density can arise already from 3+1 dimensional purely classical CGC
simulations, if incoming nuclei have a finite extent in the beam direction.
As one would expect, the Gaussian profiles become broader at higher
collision energy. In principle, we can cover the wide range from very
thin nuclei with almost boost-invariant behavior to thick nuclei at
low collision energies and narrow Gaussian profiles. We simulate the
collision in the laboratory frame, see Fig.~\ref{fig:CPIC}, which
makes it necessary to include the propagating nuclei already before
and during the collision.

The nuclei in the CGC picture consist of hard partons that are surrounded
by soft gluons. The hard partons can be described as classical color
charges moving at the speed of light, while the soft gluons form a
highly occupied coherent non-Abelian gauge field. The collision of
two such infinitely thin condensates produces the Glasma whose evolution
can be described classically by solving the Yang-Mills equations for
early proper times. At finite nuclei width, the collision region is
not pointlike anymore, and the nuclei, the collision, and the evolution
of the Glasma can not be described separately and require one consistent
simulation that covers all these steps.

A suitable  numerical method was developed in our previous publication
\citep{Gelfand:2016yho} based on the colored particle-in-cell method
(CPIC) \citep{Dumitru:2005hj,Moore:1997sn,Hu:1996sf,Strickland:2007},
which is a non-Abelian extension of the particle-in-cell method for
the simulation of Abelian plasmas \citep{Esirkepov2001,Verboncoeur2005}.
In contrast to the traditional approach of simulating the Glasma,
where the field equations are solved in the forward light-cone parametrized
by proper time $\tau$ and space-time rapidity $\eta_{s}$, we describe
the collision in the laboratory frame using the lab-frame time $t$
and the longitudinal beam direction $z$. The most striking difference
of this approach is the explicit inclusion of the nuclei in the simulation,
whereas in a boost-invariant simulation the information about the
nuclei and their color currents is completely encoded in the initial
conditions at the boundary of the light-cone, i.e.\ $\tau=0$. To
solve this problem numerically we simulate the continuous color charge
densities of the nuclei with a large number of color-charged point-like
particles, mimicking the dynamics of the continuous cloud of color
charges on a lattice.    This enables us to describe the full
3+1 dimensional collision and the subsequent evolution of the Glasma
beyond the boost-invariant approximation. For a more detailed description
we refer the reader to \citep{Gelfand:2016yho}.

The initial conditions in our simulation differ from the traditional
approach as well. Instead of starting at $\tau=0$, our simulation
begins before the collision with the nuclei well-separated in the
longitudinal direction. Here we quickly review how to solve the Yang-Mills
equations in the covariant gauge and the transformation to the temporal
gauge in the laboratory frame for a single nucleus. We base our model
of the initial state on the McLerran-Venugopalan (MV) model \citep{MV1,MV2},
extended by a thickness parameter in longitudinal direction. The transverse
charge density $\rho^{a}(x_{T})$ as a function of the transverse
coordinate $x_{T}$ is a random variable following the usual gauge-invariant
Gaussian probability functional $W[\rho]$ with the two-point correlation
function
\begin{equation}
\ev{\rho^{a}(x_{T})\rho^{b}(y_{T})}=g^{2}\mu^{2}\delta^{ab}\delta^{(2)}(x_{T}-y_{T}),
\end{equation}
where $\mu$ is the MV model parameter controlling average color charge
density and $g$ is the Yang-Mills coupling constant. For a nucleus
moving in the positive $z$ direction, we embed this two-dimensional
charge density into the three-dimensional laboratory frame via $\rho^{a}(x_{T},x^{-})=f(x^{-})\rho^{a}(x_{T})$
with a longitudinal profile function $f(x^{-})$, where $x^{\pm}\equiv\left(t\pm z\right)/\sqrt{2}$
are the usual light-cone coordinates. For the longitudinal profile
we choose a Gaussian
\begin{equation}
f(x^{-})=\frac{1}{\sqrt{2\pi}L}\exp\left(-(x^{-})^{2}/L^{2}\right),
\end{equation}
where we introduce the thickness parameter $L$. In the limit of $L\rightarrow0$
we have $f(x^{-})\propto\delta(x^{-})$ and restore the boost-invariant
limit of the original MV model. Note that this model explicitly neglects
non-trivial longitudinal color structure \citep{Fukushima:2007ki}.
The only non-vanishing component of the light-like color current of
the nucleus is then given by 
\begin{align}
J_{\text{cov}}^{+}(x_{T},x^{-}) & =\sqrt{2}f(x^{-})\rho^{a}(x_{T})t^{a},
\end{align}
where $t^{a}$ are the generators of the gauge group SU($N$). The
subscript ``$\text{cov}$'' denotes that this defines the color
current in the covariant gauge $\p_{\mu}A^{\mu,a}=0$. Using this
ansatz we can solve the Yang-Mills equations
\begin{equation}
D_{\mu}F^{\mu\nu}=J^{\nu},
\end{equation}
in the covariant gauge by finding a solution to the two-dimensional
Poisson equation 
\begin{equation}
-\Delta_{T}A^{+}(x_{T},x^{-})=J_{\text{cov}}^{+}(x_{T},x^{-}),
\end{equation}
which is solved by 
\begin{eqnarray}
\phi^{a}(x_{T}) & = & \intop_{0}^{\Lambda}\frac{d^{2}k_{T}}{\left(2\pi\right)^{2}}\frac{\tilde{\rho}^{a}(k_{T})}{k_{T}^{2}+m^{2}}e^{-ik_{T}\cdot x_{T}},\label{eq:poisson_solution}\\
A^{+}(x_{T},x^{-}) & = & \sqrt{2}f(x^{-})\phi^{a}(x_{T})t^{a},
\end{eqnarray}
where $\tilde{\rho}^{a}(k_{T})$ is the Fourier transform of $\rho^{a}(x_{T})$.
We introduced an infrared regulator $m$ and an ultraviolet cutoff
$\Lambda$, since the MV model is both infrared and UV divergent.
The regularization in (\ref{eq:poisson_solution}) should be read
as a modification of the charge densities $\rho^{a}(x_{T})$ while
the field equations remain unchanged.

Our numerical method requires the gauge fields to satisfy the temporal
gauge condition $A_{0}^{a}=0$. Switching to this gauge from the covariant
gauge renders the fields purely transverse. The transverse field components
and the color current are given by
\begin{equation}
A_{i}(x_{T},x^{-})=\frac{1}{ig}V(x_{T},x^{-})\p_{i}V^{\t}(x_{T},x^{-}),
\end{equation}
\begin{equation}
J^{+}(x_{T},x^{-})=V(x_{T},x^{-})J_{\text{cov}}^{+}(x_{T},x^{-})V^{\t}(x_{T},x^{-}),
\end{equation}
with the temporal Wilson line
\begin{align}
V(x_{T},x^{-}) & =\mathcal{T}\exp\bigg(-ig\intop_{-\infty}^{t}dt'f(x'^{-})\phi(x_{T})\bigg).
\end{align}
Having set up the initial conditions the system can be evolved forward
in time via the equations of motion on the lattice. 

 In order to fix the other parameters of the initial conditions for
a given collision energy $\sqrt{s_{NN}}$ we take the following approach:
the longitudinal thickness $L$ introduced in our implementation of
the MV model somehow needs to be related to the Lorentz factor $\gamma$.
It seems natural that $L$ should be proportional to the Lorentz contracted
length of the nucleus $\gamma^{-1}R$, where $R$ is the nuclear radius.
 One possibility is to define \citep{Gelfand:2016yho}
\begin{equation}
\gamma=\frac{R}{2L}.\label{eq:gamma_R_definition}
\end{equation}
This way $L$ is fixed by the geometry of the Lorentz-contracted nucleus.
We determine the saturation momentum $Q_{s}$ using the estimation
$Q_{s}^{2}\approx\left(\sqrt{s_{NN}}\right)^{0.25}\,\mbox{GeV}^{2}$
with $\sqrt{s_{NN}}$ given in $\mbox{GeV}$ \citep{Lappi:2006hq,Lappi:2007ku,Kharzeev:2001gp}.
The coupling constant $g$ is set by the one-loop beta function at
the saturation scale $Q_{s}$, which gives values close to $g\approx2$.
The relation between the MV model parameter $\mu$ and $Q_{s}$ is
non-trivial \citep{Lappi:2007ku} and for simplicity we choose $0.75\, g^{2}\mu\simeq Q_{s}$
as suggested in \citep{Schenke:2012fw}. The Lorentz gamma factor
is given by $\gamma=\sqrt{s_{NN}}/(2m_{N})$ with the nucleon mass
$m_{N}\approx1\:\mbox{GeV}$ and consequently we can find $L$ via
Eq.~(\ref{eq:gamma_R_definition}), setting $R$ to the radius of
a Gold nucleus. Since our results depend on the IR regulator $m$
we try out different values: $m=0.2\,\mbox{GeV}$, $m=0.4\,\mbox{GeV}$
and $m=0.8\,\mbox{GeV}$. The UV modes are regulated by $\Lambda=10\:\mbox{GeV}$.
The simulation is performed with the gauge group $\mbox{SU}(2)$ instead
of SU$(3)$, which should give reasonable qualitative results \citep{Ipp:2010uy}.

During the simulation we record the components of the energy-momentum
tensor $T^{\mu\nu}$ in the laboratory frame and average over a number
of collision events to obtain the expectation value $\ev{T^{\mu\nu}}$.
In the MV model most of the $T^{\mu\nu}$ components vanish after
averaging over all initial conditions due to homogeneity and isotropy
in the transverse plane. Therefore the energy-momentum tensor reduces
to
\begin{equation}
\ev{T^{\mu\nu}}=\left(\begin{array}{cccc}
\ev{\e} & 0 & 0 & \ev{S_{L}}\\
0 & \ev{p_{T}} & 0 & 0\\
0 & 0 & \ev{p_{T}} & 0\\
\ev{S_{L}} & 0 & 0 & \ev{p_{L}}
\end{array}\right),
\end{equation}
where $\ev{\varepsilon}$ is the energy density, $\ev{p_{L}}$ and
$\ev{p_{T}}$ are the longitudinal and transverse pressure components
and $\ev{S_{L}}$ is the longitudinal component of the Poynting vector.
The local rest frame energy density is obtained by diagonalizing the
energy-momentum tensor $\ev{T_{\,\,\nu}^{\mu}}$:

\begin{eqnarray}
\ev{\e_{\text{loc}}} & = & \frac{1}{2}\bigg(\ev{\e}-\ev{p_{L}}\nonumber \\
 &  & +\sqrt{\left(\ev{\e}+\ev{p_{L}}\right)^{2}-4\ev{S_{L}}^{2}}\bigg).\label{eq:loc_rest_frame_eng}
\end{eqnarray}
 Given the electric and magnetic fields $E_{i}^{a}$ and $B_{i}^{a}$
we have 
\begin{eqnarray}
\ev{\e} & = & \frac{1}{2}\ev{E_{T}^{2}+B_{T}^{2}+E_{L}^{2}+B_{L}^{2}},\\
\ev{p_{T}} & = & \frac{1}{2}\ev{E_{L}^{2}+B_{L}^{2}},\label{eq:pT_eL}\\
\ev{p_{L}} & = & \frac{1}{2}\ev{E_{T}^{2}+B_{T}^{2}-E_{L}^{2}-B_{L}^{2}},\\
\ev{S_{L}} & = & \ev{\big(\vec{E}^{a}\times\vec{B}^{a}\big)_{L}}.
\end{eqnarray}
Using proper time $\tau=\sqrt{t^{2}-z^{2}}$ and space-time rapidity
$\eta_{s}=\frac{1}{2}\ln\left[\left(t-z\right)/\left(t+z\right)\right]$
we can plot the profile of the energy density as a function of $\eta_{s}$
for various values of $\tau$. Due to the extended collision region
there is some ambiguity in setting the coordinate origin for the transformation
to the comoving frame. We choose the space-time coordinate of the
maximum of $\ev{p_{T}(t,z)}$, which is generated by the initially
purely longitudinal Glasma fields. This coordinate origin is slightly
later than the space-time coordinates defined by the maximum overlap
of the nuclei.

\begin{figure}
\begin{centering}
\includegraphics{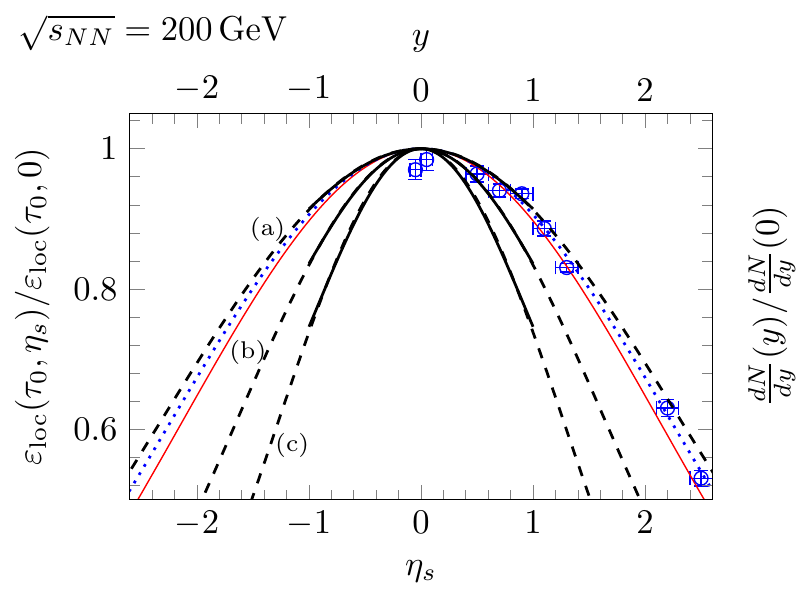}
\par\end{centering}

\protect\caption{Comparison of the space-time rapidity profile of the local rest frame
energy density $\protect\e_{\text{loc}}(\tau_{0},\eta_{s})$ for a
RHIC-like collision (thick solid lines) at $\tau_{0}=1\,\mbox{fm}$/c,
a measured profile of $\pi^{+}$ multiplicity $dN/dy$ at RHIC (data
points) and Gaussian fits (dashed and dotted lines) for our simulation
and experimental data ($\sigma_{\text{exp}}=2.25$). The value of
the infrared regulator $m$ modifies the width of the profiles: (a)
$m=0.2\,\mbox{GeV}$ with $\sigma_{\eta}=2.34$, (b) $m=0.4\,\mbox{GeV}$
with $\sigma_{\eta}=1.66$ and (c) $m=0.8\,\mbox{GeV}$ with $\sigma_{\text{\ensuremath{\eta}}}=1.28$.
Data is taken from Ref \citep{Bearden:2004yx}. The thin red line
corresponds to the profile predicted by the Landau model with $\sigma_{\text{Landau}}=\sqrt{\ln\gamma}\approx2.15$.
\label{fig:RHIC200}}
\end{figure}

In Figure \ref{fig:RHIC200} we see a calculation of the rapidity
profile of the local rest frame energy density for a RHIC-like collision:
choosing parameters $\sqrt{s_{NN}}=200\,\mbox{GeV}$ and $m=0.2\,\text{GeV}$,
we obtain an approximate Gaussian profile of the local rest frame
energy density, 
\begin{equation}
\e_{\text{loc}}(\tau_{0},\eta_{s})\approx\e_{\text{loc}}(\tau_{0},0)\exp\left(-\frac{\eta_{s}^{2}}{2\sigma_{\eta}^{2}}\right),
\end{equation}
in the range $\eta_{s}\in(-1,1)$. Using a Gaussian fit we compute
$\sigma_{\eta}$ and extrapolate to higher $\eta_{s}$. At higher
values of the IR regulator $m$ the rapidity profiles become more
narrow, which shows that the transverse size of the correlated color
structures $\sim m^{-1}$ has an effect on broken boost invariance.
In the CGC literature it is well-known that quantities such as the
initial energy density or the gluon multiplicity\textcolor{red}{}
can be sensitive to the choice of the infrared regulator \citep{Lappi:2007ku,Fukushima:2007ki,Fujii:2008km,Fukushima:2011nq}.
Here we add another example of a strong dependency on the infrared
regulation. 

Even though the results of our model should be regarded as more qualitative
than quantitative, we compare our findings to experimental results:
it is an interesting observation that the rapidity profile of the
energy density for $m=0.2\,\text{GeV}$ agrees with the measured rapidity
profile of pion multiplicities for the most central collisions at
RHIC. Of course, a direct comparison of $\e_{\text{loc}}$ and $dN/dy$
profiles is not strictly valid: the gluon number distribution can
be somewhat broader than the energy density \citep{Hirano:2004en}.
The correct approach would be to use our results as initial conditions
for hydrodynamic simulations in order to make a more direct connection
with measured observables. The subsequent hydrodynamic expansion of
the system likely increases the width of the profiles further as mentioned
in \textcolor{black}{\citep{Schenke:2016ksl}}\textcolor{red}{},
which would favor the more narrow curves (b) and (c) in Figure \ref{fig:RHIC200}.
Under these assumptions, the width of the rapidity profile of measured
charged particle multiplicities can be seen as an upper limit for
realistic rapidity profiles computed from our simulation. We also
compare to the Gaussian rapidity profile of the hydrodynamic Landau
model \citep{Landau:1953gs} with $\sigma_{\text{Landau}}=\sqrt{\ln\gamma}$.

\begin{figure}
\begin{centering}
\includegraphics{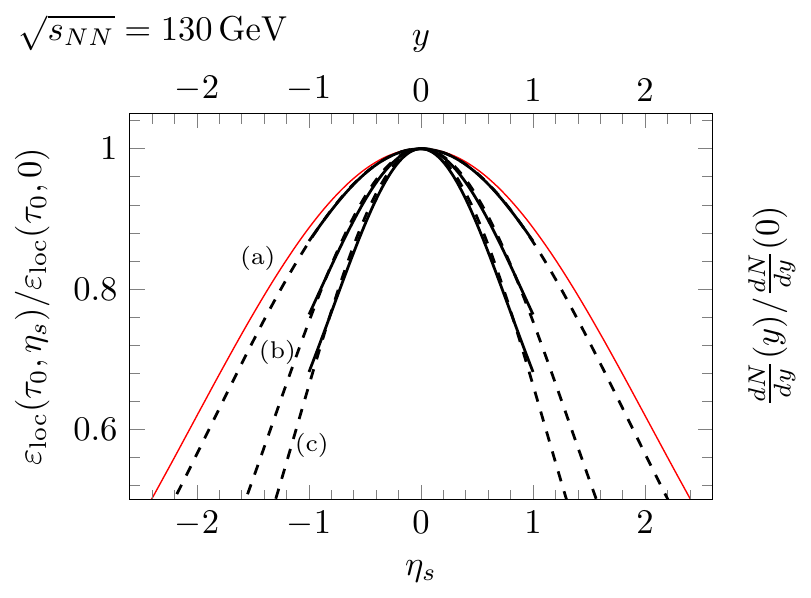}
\par\end{centering}

\protect\caption{Simulation results of collisions at $\sqrt{s_{NN}}=130\,\text{GeV}$
at $\tau_{0}=1\,\mbox{fm}$/c. The black solid lines are the computed
profiles within $\eta_{s}\in(-1,1)$. The dashed lines are fits to
Gaussian profiles: (a) $m=0.2\,\text{GeV}$ with $\sigma_{\eta}=1.87$,
(b) $m=0.4\,\text{GeV}$ with $\sigma_{\eta}=1.33$ and (c) $m=0.8\,\text{GeV}$
with $\sigma_{\eta}=1.10$. The thin red line is the result predicted
by the Landau model with $\sigma_{\text{Landau}}\approx2.04$. Compared
to Figure \ref{fig:RHIC200} the profiles are more narrow than the
Landau result. \label{fig:RHIC130}}
\end{figure}

The profiles have been computed at $\tau_{0}=1\,\mbox{fm/c}$, which
roughly corresponds to the transition from the Glasma to the QGP.
In general we observe that the rapidity profiles quickly converge
for $\tau\gtrsim0.3\,\mbox{fm/c}$ and afterwards become independent
of $\tau$. We also observe free-streaming behavior signaled by $\e_{\text{loc}}\propto1/\tau$
for a wide range of $\eta_{s}$ and longitudinal velocities $v_{z}\sim\frac{z}{t}$.
This implies that there is only negligible flow of energy between
different rapidity directions. Our results suggest that the rapidity
dependence of the Glasma is fixed early on in the collision and remains
unchanged thereafter. To see the effects of increased longitudinal
thickness, we repeat the same calculation for $\sqrt{s_{NN}}=130\,\mbox{GeV}$
in Figure \ref{fig:RHIC130}. We fit to a Gaussian shape and find
that, as expected, the profiles become more narrow as compared to
the RHIC-like case. This time, our results are also more narrow than
Landau's prediction.

\begin{figure}
\begin{centering}
\includegraphics{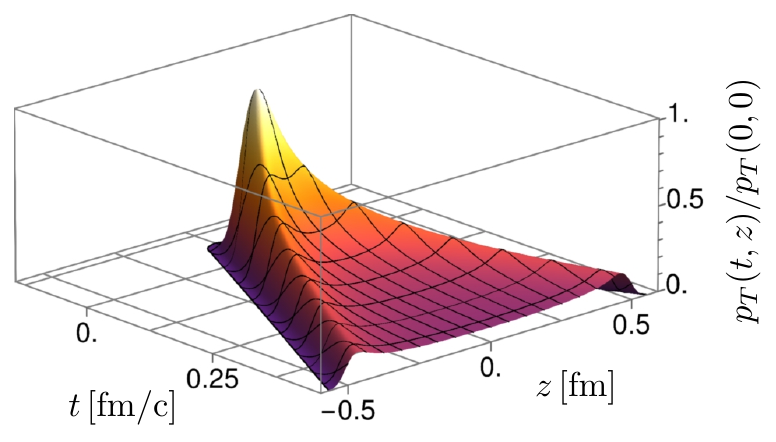}
\par\end{centering}

\protect\caption{Space-time distribution of the transverse pressure $\protect\ev{p_{T}(t,z)}$
normalized to the maximum at the coordinate origin. The same parameters
as in Figure \ref{fig:RHIC200} (a) are used. In the Glasma the transverse
pressure is generated by longitudinal magnetic and electric fields
and equivalent to the longitudinal component of the energy density
$\protect\ev{\protect\e_{L}(t,z)}$. The drop of $\protect\ev{\protect\e_{L}(t,z)}$
along the boundary of the light-cone is quite steep and becomes even
steeper when decreasing the collision energy, resulting in more narrow
rapidity profiles. \label{fig:ep_L_spacetime}}
\end{figure}

Investigating the reason behind the non-boost-invariant creation of
the Glasma we look at the space-time distribution of the transverse
pressure $\ev{p_{T}(z,t)}$ in and along the forward light-cone in
Figure \ref{fig:ep_L_spacetime}. As can be seen from Eq. (\ref{eq:pT_eL}),
the transverse pressure is solely due to the presence of longitudinal
fields, i.e.\ $\ev{p_{T}(z,t)}$ is equivalent to the energy density
generated by longitudinal fields $\ev{\e_{L}(z,t)}$. In the boost-invariant
case the longitudinal fields would be constant along the boundary
of the light-cone as determined by the boost-invariant Glasma initial
conditions at $\tau=0$. In our simulations we see very different
behavior: the longitudinal fields are mostly centered around the maximum
in the extended collision region $t\sim z\sim0$ and decrease rather
quickly along the $t=\pm z$ boundaries. Since the initially longitudinal
fields are the starting point of the evolution of the Glasma, this
sharp decrease means that for some fixed proper time $\tau$ there
is less Glasma at higher values of rapidity $\eta_{s}$, leading to
the observed Gaussian profiles of the energy density. Furthermore,
we see a mostly constant spatial distribution of $\ev{p_{T}(z,t)}$
in Figure \ref{fig:ep_L_spacetime} for later times $t$. A similar
distribution has been found in holographic models of heavy-ion collisions
\citep{Casalderrey-Solana:2013aba,vanderSchee:2015rta}.\textcolor{red}{}

The results in Figures \ref{fig:RHIC200}, \ref{fig:RHIC130} and
\ref{fig:ep_L_spacetime} have been computed on a lattice with $2048$
cells in the longitudinal direction with a length of $6\,\mbox{fm}$
and $192^{2}$ cells for the transverse plane with an area of $\left(6\,\mbox{fm}\right)^{2}$
and a statistical average over $15$ events. The tetragonal lattice
(with much smaller longitudinal than transverse lattice spacing) is
more suited for describing the collision of Lorentz-contracted nuclei.
By varying the lattice spacings, the results were checked for discretization
errors. Although accessing wider rapidity ranges is possible in principle,
we restricted the results to $\eta_{s}\in(-1,1)$ due to numerical
issues: at high space-time rapidity $\eta_{s}$ near the boundary
of the light cone, the computation of $\e_{\text{loc}}$ via Eq.~(\ref{eq:loc_rest_frame_eng})
becomes very sensitive to cancellations in the square root term. Furthermore,
the fields of the nuclei, being proportional to the longitudinal profile
$f(x^{-})$, become increasingly larger compared to the Glasma fields
as $L\rightarrow0$, which converge to the finite, boost-invariant
result at mid-rapidity for high $\sqrt{s_{NN}}$. This leads to an
unwanted modification of $\e_{\text{loc}}$ by large transverse fields
of the nuclei at larger $\eta_{s}$. We cannot strictly separate the
Glasma from the nucleus fields, which is a clear disadvantage to simulations
performed strictly in the forward light cone \citep{Schenke:2016ksl}.
For these reasons we currently can not obtain reliable results for
collision energies present at the LHC and further improvements in
the numerical scheme are needed.

Concluding, we found that Gaussian rapidity profiles in energy density
can arise from CGC collisions of nuclei with finite longitudinal thickness
in classical 3+1 dimensional Yang-Mills simulations. The width of
the profiles is controlled by the energy of the incoming nuclei, but
depends also crucially on the infrared modes of the fields. Presumably
the infrared dependencies of our results could be fixed by more realistic
initial conditions obtained from a JIMWLK evolution. It would be interesting
to understand the mechanism behind the creation of Glasma fields in
a non-boost-invariant setting with finite nucleus thickness, and in
particular also the connection to infrared regulation. In any case,
our result shows that there is a mechanism that breaks boost invariance
at the classical level in the Glasma related to the finite thickness
of the colliding nuclei, and further investigations are needed to
see how this effect compares to other previously studied approaches
like the rapidity dependence of JIMWLK. It is also exciting to see
that by lifting the assumption of boost invariance, we have shown
that our weak coupling results are in qualitative agreement with the
strong coupling results exhibiting a Gaussian rapidity profile of
the rest frame energy density and similar transverse pressure distributions
in the laboratory frame \citep{Casalderrey-Solana:2013aba}. That
is to say, our result could be interpreted to indicate that the difference
between the previous weak and strong coupling simulations is due to
the initial conditions, not the weak or strong coupling dynamics.
\textcolor{red}{}

This work has been supported by the Austrian Science Fund FWF, Project
No.~P 26582-N27 and Doctoral program No.~W1252-N27. The computational
results presented have been achieved using the Vienna Scientific Cluster.
We would like to thank Aleksi Kurkela for useful discussions. 

\bibliographystyle{apsrev4-1}
\addcontentsline{toc}{section}{\refname}\bibliography{rapidity-arxiv,rapidity-published}

\end{document}